\documentclass[a4paper,11pt,dvips]{article}
\usepackage{graphicx}
\usepackage{amssymb}
\usepackage{latexsym}
\usepackage{color}

\newcommand{\bra}{\begin{array}}
\newcommand{\era}{\end{array}}
\newcommand{\beq}{\begin{equation}}
\newcommand{\eeq}{\end{equation}}
\newcommand{\bqr}{\begin{eqnarray}}
\newcommand{\eqr}{\end{eqnarray}}

\def\BC{\bb C}
\def\_\BC{\bbi C}

\def\( {\left(}
   \def\) {\right)}
\def\[ {\left[}
\def\] {\right]}
\def\no2 {{\textstyle{n\over 2}}}

\def\dag {{\dagger}}

\newcommand{\si}{\sigma}

\newcommand{\pa}{\partial}
\newcommand{\al}{\alpha}

\newcommand{\lga}{\longrightarrow}

\newcommand{\lb}{\label}
\newcommand{\non}{\nonumber}

\begin{document}
\begin{titlepage}
\setcounter{page}{1}
\renewcommand{\thefootnote}{\fnsymbol{footnote}}

\begin{flushright}
\end{flushright}

\vspace{5mm}
\begin{center}

{\Large \bf {Transport Properties through  Double Barrier\\ Structure in Graphene}}

\vspace{5mm}

{\bf A. Jellal\footnote{\sf ajellal@ictp.it -- jellal.a@ucd.ac.ma}}$^{a,b,c}$,
{\bf E.B. Choubabi}$^{b,c}$,
{\bf H. Bahlouli}$^{d}$
and {\bf A. Aljaafari}$^{a}$

\vspace{5mm}

{$^a$\em Physics Department, College of Sciences, King Faisal University,\\
PO Box 9149, Alahsa 31982, Saudi Arabia}

{$^b$\em Saudi Center for Theoretical Physics, Dhahran, Saudi Arabia}

{$^{c}$\em Theoretical Physics Group,  
Faculty of Sciences, Choua\"ib Doukkali University},\\
{\em PO Box 20, 24000 El Jadida,
Morocco}

{$^d$\em Physics Department,  King Fahd University
of Petroleum $\&$ Minerals,\\
Dhahran 31261, Saudi Arabia}



\vspace{3cm}

\begin{abstract}

The mode-dependent transmission of relativistic ballistic massless Dirac fermion through a
graphene based double barrier structure is being investigated for various barrier parameters.
We compare our results with already published work and point out the relevance
of these findings to a systematic study of the transport properties
in double barrier structures.
An interesting situation arises when we set the potential in the leads to zero, then our 2D problem reduces
effectively to a 1D massive Dirac equation with an effective mass proportional to the quantized wave number along  the transverse direction.
Furthermore we have shown that the minimal conductivity
and maximal Fano factor remain insensitive to the ratio between the two
potentials $\left(\frac{V_2}{V_1}=\al\right)$.
\vspace{3cm}



\end{abstract}
\end{center}
\end{titlepage}


\section{Introduction}

The physics of graphene material was celebrated by attributing the 2010 physics Nobel prize
to two physicists, Novoselov and Geim, for their pioneering work on graphene. This material
was described by some physicists as "heavenly" due to its marvelous physical and transport
properties \cite{1}. The story started about seven years ago when the possibility to isolate and
investigate graphene \cite{1,2,3}, i.e. individual layers of graphite only one-atom-thick, have been
demonstrated. Further, experiments were reported showing that charge carriers in graphene behave
as two-dimensional relativistic particles with zero effective mass. In fact, one of the most
interesting aspects of the graphene problem is that its low-energy excitations are massless, chiral,
Dirac fermions. The quasiparticle excitations around the Dirac point obey a linear type of energy
dispersion law. This particular dispersion, that is only valid at low energies, mimics the physics
of quantum electrodynamics (QED) for massless fermions except for the fact that in graphene the Dirac
fermions move with a Fermi speed $v_{\sf F}$, which is 300 times smaller than the speed of light $c$. Hence,
many of the unusual properties of QED can show up in graphene but at much smaller speeds \cite{4,5,6}.
In addition Dirac fermions behave in unusual ways when compared to ordinary electrons if subjected
to magnetic fields, leading to new physical phenomena \cite{7,8} such as the anomalous quantum Hall effect,
which was observed experimentally \cite{2,3}.

On the other hand, many efforts have been employed to understand the scattering behavior of fermions
in graphene, 
among them we cite for instance \cite{11,peeters1,peeters2}.
In particular 
the authors in \cite{11}
calculated the mode-dependent transmission probability of massless Dirac fermions through an
ideal strip of graphene (length  $l$, width { $w$, in absence of }impurities or defects), to obtain the conductance
and shot noise as a function of the Fermi energy. They  found that the minimum conductance of order
$e^2/h$ at the Dirac point (when the electron and hole excitations are degenerate) is associated with
a maximum of the Fano factor (the ratio of noise power and mean current). For short and wide
graphene strips the Fano factor at the Dirac point equals $1/3$, three times smaller than for a Poisson
process. This is the same value as for a disordered metal, which is counter intuitive since the
dynamics of the Dirac fermions in graphene is ballistic.

 {To generalize the analytical approach developed in~\cite{11} and study
 other type of scattering, we investigate the behavior of massless Dirac fermions
 in a flat chip of graphene. This will be
 based on the 2D massless Dirac equation which constitutes a good description of the low energy excitations
 of the original honeycomb lattice \cite{9,10,10'}.
 The flat chip is under the influence of a short range} double barrier potential with infinite mass boundary
 condition \cite{12}, which then result in a quantization of the wave number associated with the confining $y$-direction.
 We consider the transmission of massless Dirac fermions through such a double barrier structure for various barrier parameters.
 Continuity of the wavefunction at each interface along with the infinite mass boundary condition in the $y$-direction resulted
 in  a system of eight algebraic equations for eight unknown coefficients. 
 The detailed study of such 
 system gave rise to
 a variety of interesting situations that were investigated. In particular the single barrier results~\cite{11}
 were obtained as a particular case in our model.
 Then we studied the effect of different potential parameters on the transmission, conductance and shot noise.

 An interesting situation
 arises when we set the potential floor in the leads to zero, then our 2D problem reduces effectively to a 1D massive Dirac equation with
 an effective mass proportional to the quantized wave number along the transverse direction called the  $y$-direction. Thus confinement along the $y$-direction generated
 effective masses for our fermions, which depend on the quantized wave number and its energy line spacing is proportional to the inverse
 of its width. On the other hand, we conclude that it is fairly interesting that
 the minimal conductivity and maximal Fano factor remain the same independently of the  ratio between
 the two potentials barriers involved.

The present paper is organized as follows, in section $2$, we solve the Dirac equation to
derive the energy spectrum using the infinite mass boundary condition along $y$-direction. These quantized wave numbers will serve as channel labels in dealing with tunneling
in section $3$ where the reflection and transmission amplitudes are determined through
the electric current
density.  We then explore the obtained results in section $4$ to discuss different situations
related to the characteristics of the barriers. We have also devised a parameter which correlates the barrier heights and depth so as to enable us to investigate different potential configurations.
The conductance and the Fano factor, for the present system, will be analyzed in section 5.
  In section 6, we show that the results obtained in~\cite{11} are a special case of our findings. We then conclude our work in the final section.

\section{ Theoretical model}

In the system made of graphene, the
two Fermi points, each with a two-fold band degeneracy, can be described by a low-energy continuum
approximation with a four-component envelope wavefunction whose components are labeled by a
Fermi-point pseudospin = $\pm 1$ and a sublattice forming a honeycomb. Specifically, the Hamiltonian
for one-pseudospin component 
in the vicinity of the $K$ point and in the presence of a {scalar} potential $V(x)$ can  be described by 
\begin{equation}
H = v_{\sf F}\ \vec{\sigma} \cdot \vec{p}+ V(x) {{\mathbb I}_{2}}
\end{equation}
where the pseudospin matrices $\vec\sigma $ are represented by the Pauli matrices,
$\vec{p}=-i\hbar \, \vec{\nabla }$ and $v_{\sf F} = 3ta/(2\hbar )\approx 10^{6} ms^{-1} $
is the Fermi velocity of the massless Dirac fermions, $t$ being the nearest neighbor
hopping matrix element and $a$ the carbon-carbon interatomic distance. {
The above equation can be written explicitly in matrix form using the system unit $(\hbar=c=e=v_{\sf F}=1)$}
\beq
H=  \left(%
\begin{array}{cc}
  V(x) & -i\pa_x-\pa_y \\
  -i\pa_x+\pa_y & V(x)  \\
\end{array}%
\right).
\eeq

To go further, we fix the potential $V(x)$ by considering a double barrier potential described pictorially
in Figure 1:\\

 \begin{center}
\includegraphics [width=10.10cm,keepaspectratio]
{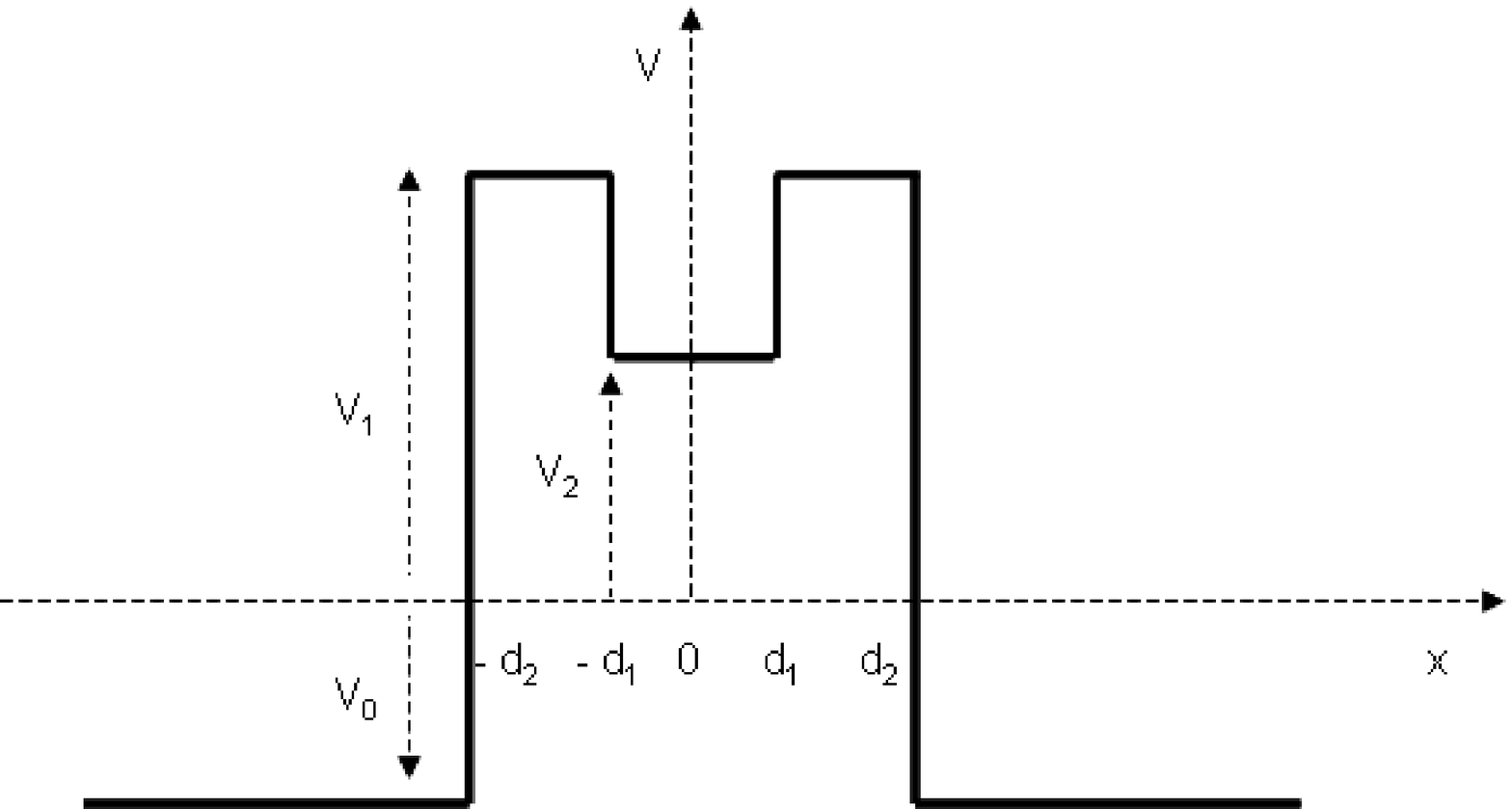}
\end{center}
\begin{center}
{\sf{Figure 1: The double barrier potential along $x$-direction}.}
\end{center}
This potential 
is defined by
\begin{equation} \label{GrindEQ__2_}
V(x)=\left\{
\begin{array}{cc} {V_{0} },  & {\left|x\right|\ge d_{2} } \\
{V_{1} },  & \qquad \ {d_{1} <\left|x\right|<d_{2} } \\
{V_{2} }, & {\left|x\right|\le d_{1} }
\end{array}\right.
\end{equation}
where $V_{1,2} $ and $d_{1,2} $ are positive potential parameters (see Figure 1) characterizing the double barrier structure.
We divide the configuration space into five regions numbered $j = 1, \cdots 5$ associated with piece-wise constant potential sections.
In the outside barrier regions $V=V_{0} $, $V=V_{1} $ for $j = 2,4$ in the barrier regions and finally $V=V_{2} $ for $j = 3$
in the well region. Each region $j$ is characterized by its wave vector $\vec
{k_{j} }=k_{jx} \vec i+k_{jy} \vec j$ such that $k_{j}^{2} =k_{jx}^{2} +k_{jy}^{2} $. The absence of potential along
the $y$-direction leads to the conservation of  $k_{jy} $ in all regions
and  its value is subsequently determined by
the infinite mass boundary condition.

Solving the eigenvalue equation
\begin{equation} \label{GrindEQ__1_}
{  \left(%
\begin{array}{cc}
  V(x) & -i\pa_x-\pa_y \\
  -i\pa_x+\pa_y & V(x)  \\
\end{array}%
\right)
\psi =E\psi} 
\end{equation}
we get plane wave solutions in all constant potential regions, such that in the $j$-th region we can write 
\begin{equation} \label{GrindEQ__3_}
\psi _{j}(x,y) =\frac{1}{\sqrt{2} }
\left(\begin{array}{c} {1} \\ {z_{j} }
\end{array}\right)
e^{{i}\vec{{ k}}_{j} \cdot \vec{r}}
\end{equation}
where the parameter $z_j$ is given by
\beq
 z_{j} =s_{j} \frac{k_{jx} +i\, k_{jy} }
{\sqrt{k_{jx}^{2} +k_{jy}^{2} } } =s_{j} e^{{i}\theta _{j} }
\eeq
with the sign function $s_{j} ={\sf sign}\left(E-V_{j} \right)$,  the phase $\theta _{j} =\arctan\left(\frac{k_{jy} }{k_{jx} }\right )$
and the propagation wave number {$k_{jx} =\sqrt{(E-V_{j} )^{2}  -k_{jy}^{2} } $}.
The corresponding energy eigenvalue reads as
\begin{equation} \label{GrindEQ__4_}
{E=V_{j} + s_{j}  \sqrt{k_{jx}^{2} +k_{jy}^{2} }}
\end{equation}
where the $s_{j} =\pm 1$ sign distinguishes electrons from hole conduction in the $j$-th constant potential region.
One can show that the complex numbers $z_j$ satisfies the identity
\begin{equation}\lb{zj2}
{1+z_{j}^{2}=2s_{j} \frac{k_{jx}z_{j}}{\left|E-V_{j}\right|}}
\end{equation}
which will play a crucial role in analyzing the tunneling effect
and, in particular, in discussing different potential configurations used. 

One way to confine the particle along the transverse $y$-direction is to use a scalar potential
(vector potential will not serve the purpose due the Klein tunneling)  that rises to very large
values at the boundary, the width of the strip along $y$-direction is denoted by $w$. This is achieved
by using a mass term in the Dirac equation that rises to infinity at the edges $y = 0$ and $y = w$. Otherwise,
the infinite mass boundary condition gives us
the quantification of the transverse wave vector $k_{jy}$. To determine this tangential component, we will
consider a four-component eigenspinor, denoted by $\psi\equiv\psi_{n,k_{j}}(x,y)=\left(%
  \psi_{1},  \psi_{2},  \psi_{3},  \psi_{4}
\right)^{t}$. The first two components satisfy the Dirac equation
without the mass term and the second two components satisfy the
same equation by mapping $p_{jy} \rightarrow -p_{jy}$.
For simplicity, we use the notations $k_{jy}\equiv q_{n}$ and $k_{jx} \equiv k_{j}$
to write
the four-component eigenspinor as
\begin{equation}
\psi
=a_{n}\left(%
\begin{array}{c}
  1\\
  z_{j}\\
  0\\
  0 \\
\end{array}%
\right)e^{i(q_{n}y+k_{j}x)}+a'_{n}\left(%
\begin{array}{c}
  0\\
  0\\
  z_{j}\\
  1 \\
\end{array}%
\right)e^{i(q_{n}y+ k_{j}x)}+b_{n}\left(%
\begin{array}{c}
 z_{j}\\
 1\\
  0\\
  0 \\
\end{array}%
\right)e^{-i(q_{n}y - k_{j}x)}+b'_{n}\left(%
\begin{array}{c}
  0\\
  0 \\
  1\\
  z_{j}\\
\end{array}%
\right)e^{-i(q_{n}y - k_{j}x)}.
\end{equation}
The infinite mass boundary condition is expressed as follows \cite{12}
\begin{equation}
\psi |_{y=0}=\left(%
\begin{array}{cc}
  \sigma_{x} & 0 \\
  0 & - \sigma_{x} \\
\end{array}%
\right)\psi |_{y=0}, \qquad
\psi |_{y=\omega}=\left(%
\begin{array}{cc}
  -\sigma_{x} & 0 \\
  0 &  \sigma_{x} \\
\end{array}%
\right)\psi |_{y=\emph{w}}.
\end{equation}
After some algebras, we end up with
\beq
q_{n} =\frac{\pi}{{w}} \left(n+\frac{1}{2} \right)
\eeq
where $n$ is a positive integer number, which from now on denotes the propagation mode.
Thereby confining the particle along the $y$-direction leads to a quantized value of the transverse wave vector.
Finally, the energy spectrum is now given by
\begin{equation} \label{GrindEQ__400_}
{E= V_{j} + s_{j} \ \sqrt{\frac{\pi^2 }{{ w^2}} \left(n+\frac{1}{2} \right)^2 + k_{jx}^{2} }.}
\end{equation}
These results will be used to discuss different issues related to scattering phenomena, in particular,
in the next section we will investigate the reflection and transmission coefficients
through the double barrier structure.

\section{Transmission and reflection amplitudes}

The impact of the incident wave at a junction gives rise to reflected
and transmitted waves. For the reflected wave, the
wave vector $k_j$ along $x$-direction is opposite to that of the incident wave
and the corresponding  $\theta _{j}$ angle is transformed into $\pi - \theta_{j}$.
This allows us to write the spinors as
\bqr
\psi^{+}_{(j,n)}(x,y) &=& \frac{1}{\sqrt{2}}\left(
\begin{array}{c}
1 \\
 z_{j}\end{array}\right)e^{i(k_{j} x +q_{n} y)}\\
\psi^{-}_{(j,n)}(x,y)&=&\frac{1}{\sqrt{2}}\left(
\begin{array}{c}
1 \\
 -z^{-1}_{j}\end{array}\right)e^{i(-k_{j} x +q_{n} y)}
\eqr
where the superscripts are obviously associated with waves traveling to the right $(+)$ or to the left $(-)$ along the propagation direction $(x)$.
The subscripts, on the other hand, denote the scattering region $(j)$ and the propagation mode $(n)$.
In the forthcoming analysis, we will see how these can be used to explicitly determine the reflection and
transmission amplitudes.

It is straightforward to solve the tunneling problem for Dirac fermions
through the double barrier potential. Indeed, let us assume that the incident wave propagates at an
angle $\theta _{1}$ with respect to the positive $x$-direction and the impact of the incident wave on any junction gives rise
to a reflected wave and transmitted wave. In this case,
the upper and lower components of the Dirac spinor, denoted by $\varphi _{1} $ and $\varphi _{2} $,
are given  in each region by the following form
\bqr
\varphi _{1} &=& \left\{
\begin{array}{ccc}
{(e^{{ i}k_{1} x} +re^{-{ i}k_{1} x} )e^{{ i}q_{n} y} }, & & {x<-d_{2} } \\
{(ae^{{ i}k_{2} x} +be^{-{ i}k_{2} x} )e^{{ i}q_{n} y} },  &  & \qquad \ \ {-d_{2} <x<-d_{1} } \\
{(ce^{{ i}k_{3} x} +de^{-{ i}k_{3} x} )e^{{ i}q_{n} y} },  &  & \qquad \ \ {-d_{1} <x<d_{1} } \\
{(ee^{{ i}k_{2} x} +fe^{-{ i}k_{2} x} )e^{{ i}q_{n} y} },  &  & \qquad {d_{1} <x<d_{2} } \\
{te^{{ i}k_{1} x} e^{{\ i}q_{n} y} }, &  & {x>d_{2} } \\
\end{array}\right. \label{GrindEQ__5_}\\
\varphi _{2} &=&\left\{
\begin{array}{ccc} {(z_{1} e^{{ i}k_{1} x} -rz^{-1}_{1} e^{-{ i}k_{1} x} )e^{{ i}q_{n} y} }, &  & {x<-d_{2} } \\
{(az_{2} e^{{ i}k_{2} x} -bz^{-1}_{2} e^{-{ i}k_{2} x} )e^{{ i}q_{n} y} }, &  & \qquad \ \ {-d_{2} <x<-d_{1} } \\
{(cz_{3} e^{{ i}k_{3} x} -dz^{-1}_{3} e^{-{ i}k_{3} x} )e^{{ i}q_{n} y} }, &  & \qquad \ \ {-d_{1} <x<d_{1} } \\
{(ez_{2} e^{{ i}k_{2} x} -fz^{-1}_{2} e^{-{ i}k_{2} x} )e^{{ i}q_{n} y} }, & & \qquad {d_{1} <x<d_{2} } \\
{tz_{1} e^{{ i}k_{1} x} e^{{ i}q_{n} y} }, &  & {x>d_{2} }.
\end{array}\right. \label{GrindEQ__6_}
\eqr
Using the continuity of the spinor wavefunctions at the four potential discontinuities,
one arrives at a system of eight algebraic equations for the eight unknown
coefficients $(a, b, c, d, e, f, r, t)$. {These relationships can be expressed in terms of transfer matrices between different regions but this sophisticated technique is not very much needed in our case}.
Solving this system of equations for the transmission amplitude, which is of interest to us, gives
\begin{equation} \label{GrindEQ__7_}
t=\frac{A}{B+C+D+E+F}
\end{equation}
where the involved parameters read as
\begin{eqnarray} \label{GrindEQ__8_}
A &= &e^{2{ i}[(k_{2} +k_{3} )d_{1} -(k_{1} -k_{2} )d_{2} ]} (1+z_{1}^{2} )(1+z_{2}^{2} )^{2} (1+z_{3}^{2} ) \nonumber\\
B&=&e^{4{ i}k_{2} d_{2} } (z_{1} -z_{2} )^{2} (z_{2} -z_{3} )^{2} \nonumber \\
C &=& e^{4{ i}(k_{2} +k_{3} )d_{1} } (1+z_{1} z_{2} )^{2} (z_{2} -z_{3} )^{2} \nonumber \\
D &=& e^{4{ i}(k_{3} d_{1} +k_{2} d_{2} )} (z_{1} -z_{2} )^{2} (1+z_{2} z_{3} )^{2}  \\
E &=& e^{4{ i}k_{2} d_{1} } (1+z_{1} z_{2} )^{2} (1+z_{2} z_{3} )^{2} \nonumber \\
F& =&2e^{2{ i}k_{2} (d_{1} +d_{2} )} (-1+e^{4{ i}k_{3} d_{1} } )(z_{1} -z_{2} )(z_{2} -z_{3} )(1+z_{1} z_{2} )(1+z_{2} z_{3} ).\nonumber 
\end{eqnarray}

To evaluate the reflection and transmission coefficients, we introduce the electric current density $J$
for our system. After calculation, we obtain
\begin{equation} \label{GrindEQ__9_}
{\vec J=\pm { i}\psi ^{\dag } \vec\sigma \psi}
\end{equation}
where $\psi$ stands for $\psi^{\sf in}=\psi^{+}_{(1,n)}(x,y)$, $\psi^{\sf ref}=r\psi^{-}_{(1,n)}(x,y)$
and $\psi^{\sf tr}= t\psi^{+}_{(5,n)}(x,y)$.
Computing explicitly (\ref{GrindEQ__9_}) gives for the incident, reflected and transmitted current density components
{\begin{eqnarray} \label{GrindEQ__10_}
J_{x}^{\sf in} &=&  \pm { i}  (z_{1} +z_{1}^{*}) = \pm 2{i}  s_{1}\ \frac{k_{1}} {\sqrt{k_{1}^{2} +k_{n}^{2}}}  \nonumber\\
J_{x}^{\sf ref} &=& \mp {i} r^{*} r(z_{1} +z_{1}^{*} ) =\mp 2{i} r^{*} r s_{1}\ \frac{k_{1} }{\sqrt{k_{1}^{2} +q_{n}^{2} } }  \\
J_{x}^{\sf tr} &=& \pm { i} t^{*} t(z_{5} +z_{5}^{*} )=\pm 2{ i} t^{*} t s_{5}\ \frac{k_{5} }{\sqrt{k_{5}^{2} +q_{n}^{2} } } . \non
\end{eqnarray}}
The transmission and reflection coefficients, are expressed as follows
\begin{eqnarray} \label{GrindEQ__11_}
T &=&\frac{\left|J_{x}^{\sf tr} \right|}{\left|J_{x}^{\sf in} \right|}
=\left|\frac{k_{5} }{k_{1} } \right|\frac{\sqrt{k_{1}^{2} +q_{n}^{2} } }{\sqrt{k_{5}^{2} +q_{n}^{2} } } \left|t\right|^{2}
=\left|K\right||t|^{2}\\
 R &=&\frac{\left|J_{x}^{\sf ref} \right|}{\left|J_{x}^{\sf in} \right|} =|r|^{2}.
\end{eqnarray}
In our case due to the symmetry of the potential configuration in the incident and transmission regions
we have $\left|K\right|=1$, i.e. $k_1=k_5$, and  therefore we have $T=|t|^{2} $.
The above results  will be investigated numerically for different potential configurations to enable
us to extract more conclusions regarding the basic features of our system.

\section{Limiting cases and discussions}

Similarly to the single barrier problem analyzed in~\cite{kat}, let us
investigate the energy spectrum structures to understand further our system.
Recall that, the spectrum of Dirac fermions in single-layer
graphene is linear at low Fermi energies. Applying a
potential barrier $V_{j}$ in region $j$ causes a displacement of the
spectrum by an offset of $V_{j}$ as shown in Figure 2. Each conical
spectrum of graphene is the result of intersection of the energy
bands originating from sublattices $A$ (shown in red) and $B$ (shown
in blue). The portion of the spectrum, which is above the offset,
represents the conduction band and  the lower one represents the valence band.\\

\begin{center}
\includegraphics [width=8cm,keepaspectratio]
{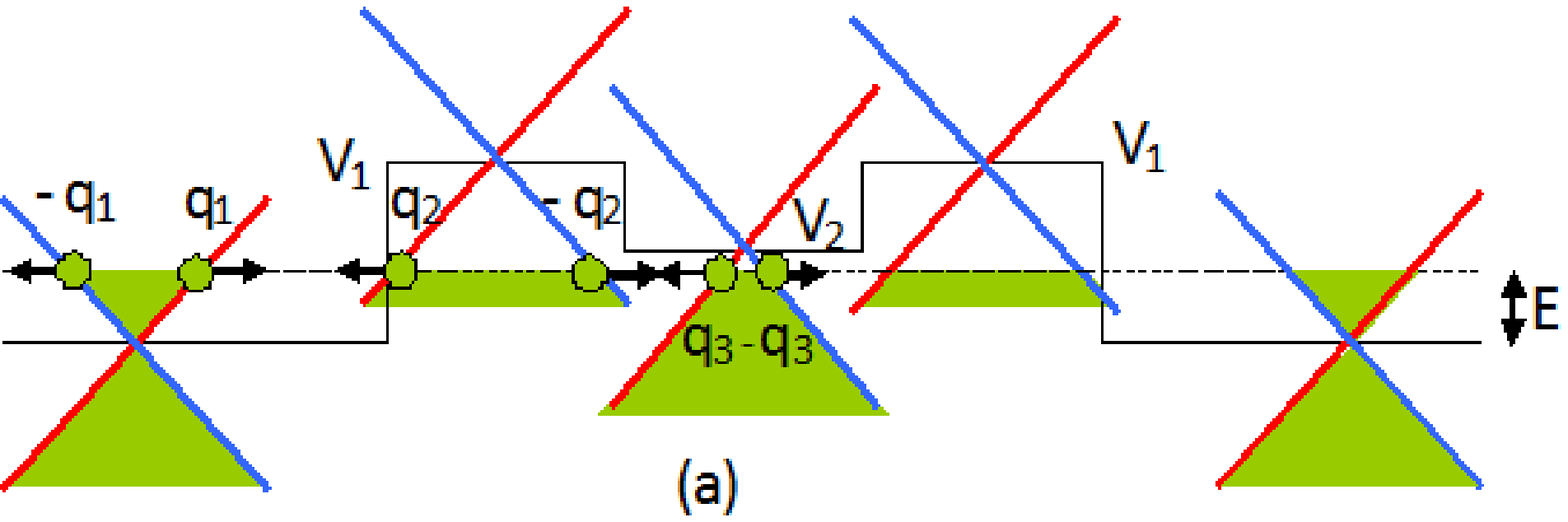}
~~~~\includegraphics [width=8cm,keepaspectratio]
{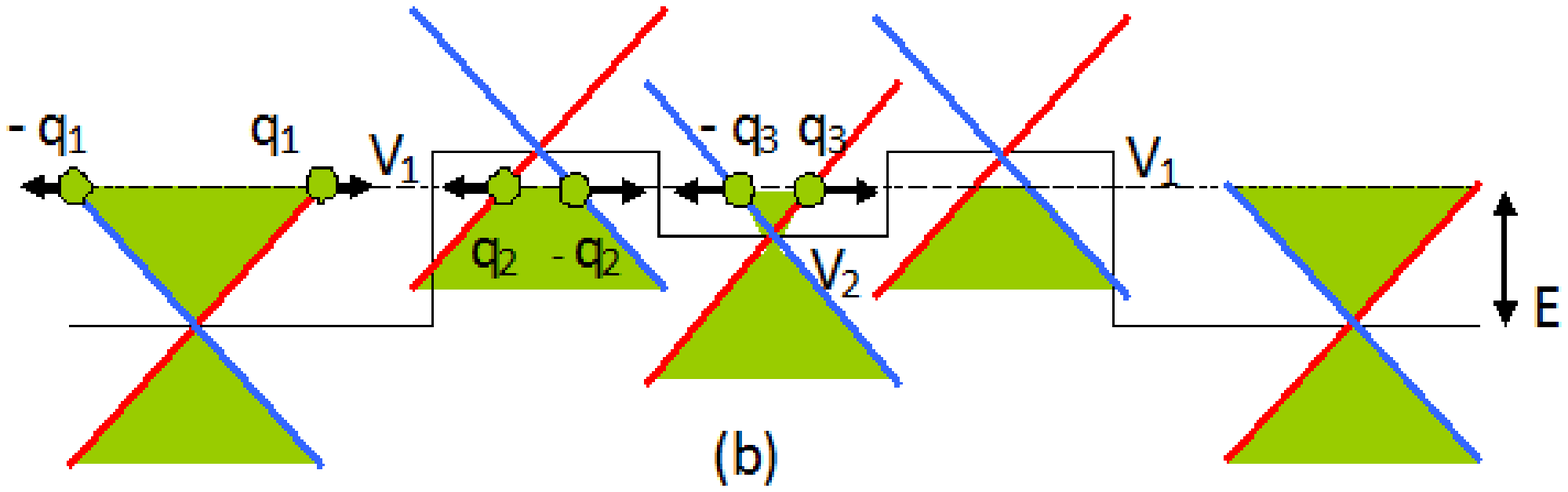}\\
\end{center}
\begin{center}
\includegraphics [width=8cm,keepaspectratio]
{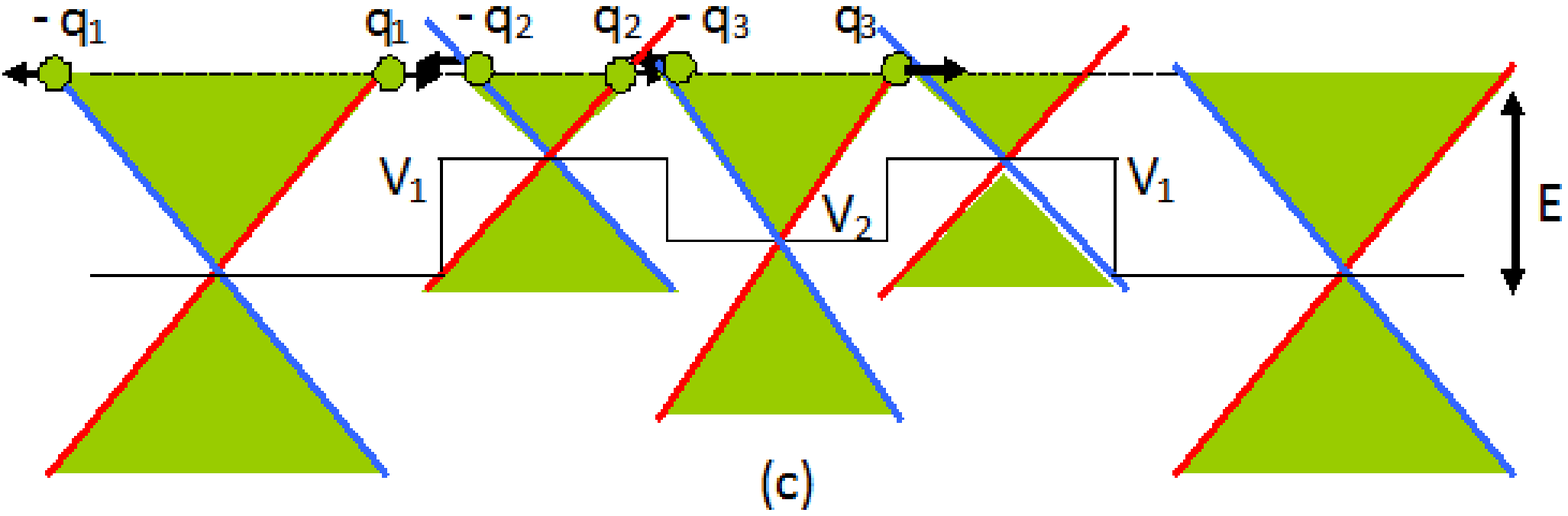}
\end{center}
{\sf{Figure 2:  Tunneling through double barriers $V_j$ in
graphene where
 (a): $V_{0}\leq E \leq V_{2} $, (b): $V_{2}\leq E \leq V_{1} $, (c): $ E\geq V_{1}$.
 The green filling indicates occupied states and  the pseudospin $\si$  is parallel
 (antiparallel) to the direction of motion of electrons (holes).}}\\

\noindent
The value of energy $E$ compared to potential barrier strength $V_{j}$ fixes the value of signature $s_{j}=\pm 1$ in  each region $j$.
This sign plays an important role because it determines the nature of the particles (electrons or holes) involved in a given region
$j$ as well as the direction of wave vectors. Clearly, the symmetry between the conduction and valence bands allows us to take $E\geq 0$
and the other case can be treated in similar way.

We are now in a position to consider solutions corresponding to different physical potential parameters
and find the associated transmission coefficients. First we start by treating the situation that
 generalizes \cite{11} by considering $V_1 = V$ and $V_2 = \alpha $ V,  $\alpha $ being
 a constant parameter.
The above potential parametrization will enable us to consider a variety of interesting physical situations
and therefore simplify the analysis of the transmission (\ref{GrindEQ__7_}). First let us consider the case  $V_{0}=0$ and 
$\alpha=1$, which represents a single barrier potential problem. In this case
(\ref{GrindEQ__7_}) becomes
\begin{equation}
t_{n}=\frac{e^{-{{i}}d(k_{1}-k_{2})}(1+z_{1}^{2})(1+z_{2}^{2})}{e^{2{{i}}dk_{2}}(z_{1}-z_{2})^{2}+(1+z_{1}z_{2})^{2}}
\end{equation}
where $d=2d_{2}=4d_{1}$.
Here, tunneling  exists only for some allowed
energies, which  depend essentially on the quantized wave vector
$q_{n}$ along $y$-direction. In Figure 3 we plot the transmission coefficient as a function of energy
for different values of the potential parameters.\\

\begin{center}
\includegraphics [width=8cm,keepaspectratio]
{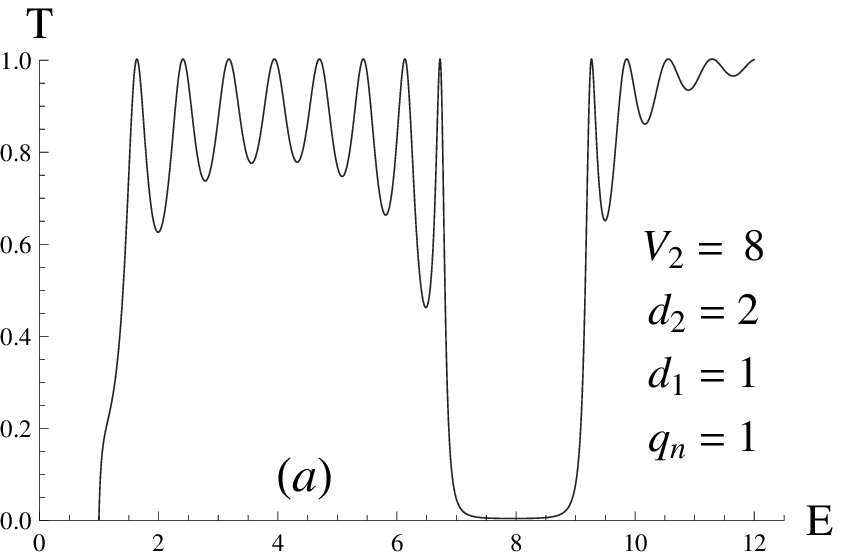}
~~~~~~\includegraphics [width=8cm,keepaspectratio]
{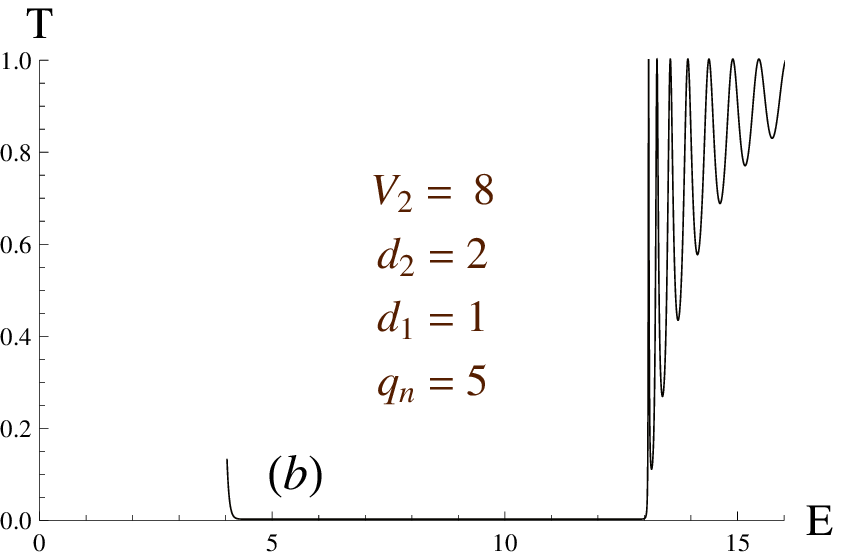}
\end{center}
\begin{center}
\includegraphics [width=8cm,keepaspectratio]
{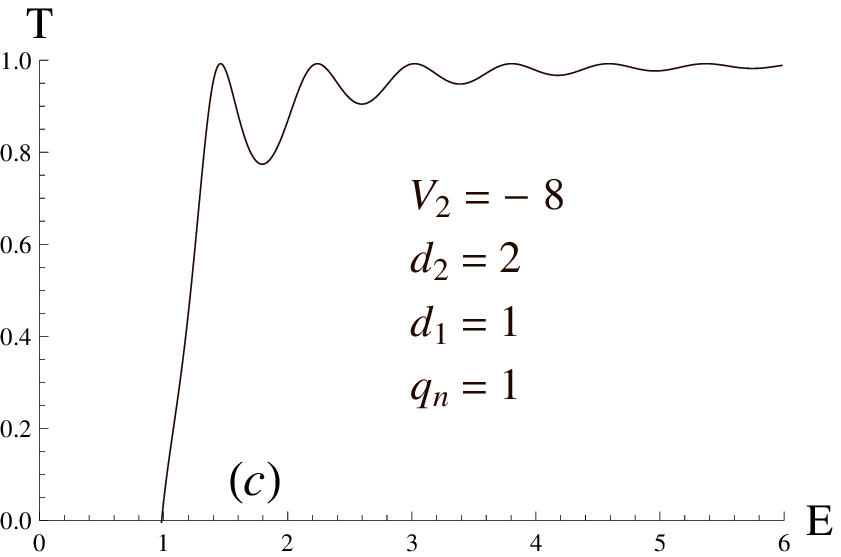}
\end{center}
{\sf{Figure 3: The transmission coefficient as a function of energy 
for three cases (a): $q_{n}\leq \frac{V}{2}$, (b):
$q_{n}\geq \frac{V}{2}$, (c): $V\leq 0${, with $V=V_2, V_0=0, \al=1$.}}}\\

\noindent Clearly, in Figure 3a, one can notice that there are four energy zones
that characterize the transmission coefficient:
\begin{itemize}
\item The first zone is bounded by the energy interval $0\leq E \leq
q_{n}$, which can be seen as  a forbidden zone.
\item The second is the Klein zone{, i.e.
 $q_{n}\leq E\leq V-q_{n}$
where $V$ is the step/barrier height and $q_n$ the dynamical particle mass,
which contains oscillations (resonances) in the transmission coefficient. This is the situation in which only
oscillatory solutions exist throughout and where the so called Klein paradox reigns \cite{Klein29}. The physical essence of the Klein tunneling lies in the prediction that according to the Dirac equation, fermions can pass through strong repulsive potentials without the exponential damping expected in usual quantum tunneling processes. The energy range in which this oscillatory behavior persists is called the Klein zone}
\item The third
zone $V-q_{n}\leq E \leq V+q_{n}$ is a bowl (window)  of zero transmission.
\item Finally, the fourth zone $E \geq
V+ q_{n}$  contains the usual high energy barrier oscillations  and asymptotically goes to unity at high energy.
\end{itemize}
On the other hand, compared to our previous work \cite{13},
we found a strong correlation between the present 2D massless Dirac
fermions and massive 1D Dirac fermions. That is,
our 2D system with $m=0$ is equivalent to 1D system with an effective mass
$m^{*}=q_{n}$. Our effective 1D system behaves as if its carriers have a dynamical mass that depends on the transverse
quantized wave number.
We can say that our effective 1D system has many carriers each associated with a quantized effective mass depending on
the transverse propagation mode $n$.
Furthermore, according to  Figure 3a and 3b one can
draw interesting conclusions. Indeed,
if $m^{*}\leq \frac{V}{2}$ then the transmission is zero for $E\leq m^{*}$
while for $m^{*}\geq \frac{V}{2}$ the transmission vanishes $E\leq V+ m^{*}$.
{In addition, the isolated {peak} $(T\ll 1)$ in Figure 3b is resulted from the fact that
the allowed energy started from the value $E=\frac{V}{2}$ and specifically it appeared at
the value $E=4$ which corresponds to $V=8$.}
However if the potential $V$ is negative, we end up with a potential
well behavior where the corresponding transmission is plotted
in Figure 3c.
Note that for any pair of $(m^{*},V)$ the allowed energy verifies
the condition $E\geq m^{*}$. In summary we can claim that the infinite mass
boundary condition generated
a dynamical mass for our original massless system, which is equivalent to space compactification of graphene
from 2D to 1D \cite{13}.

 Let us treat the double barrier case where $V_{0} = 0$ and $\alpha \neq 1$, i.e.
 $V_2< V_1$ and $V_2 > V_1$. In both cases, the transmission coefficient
 is plotted in Figure 4:\\

\begin{center}
\includegraphics [width=7.5cm,keepaspectratio]
{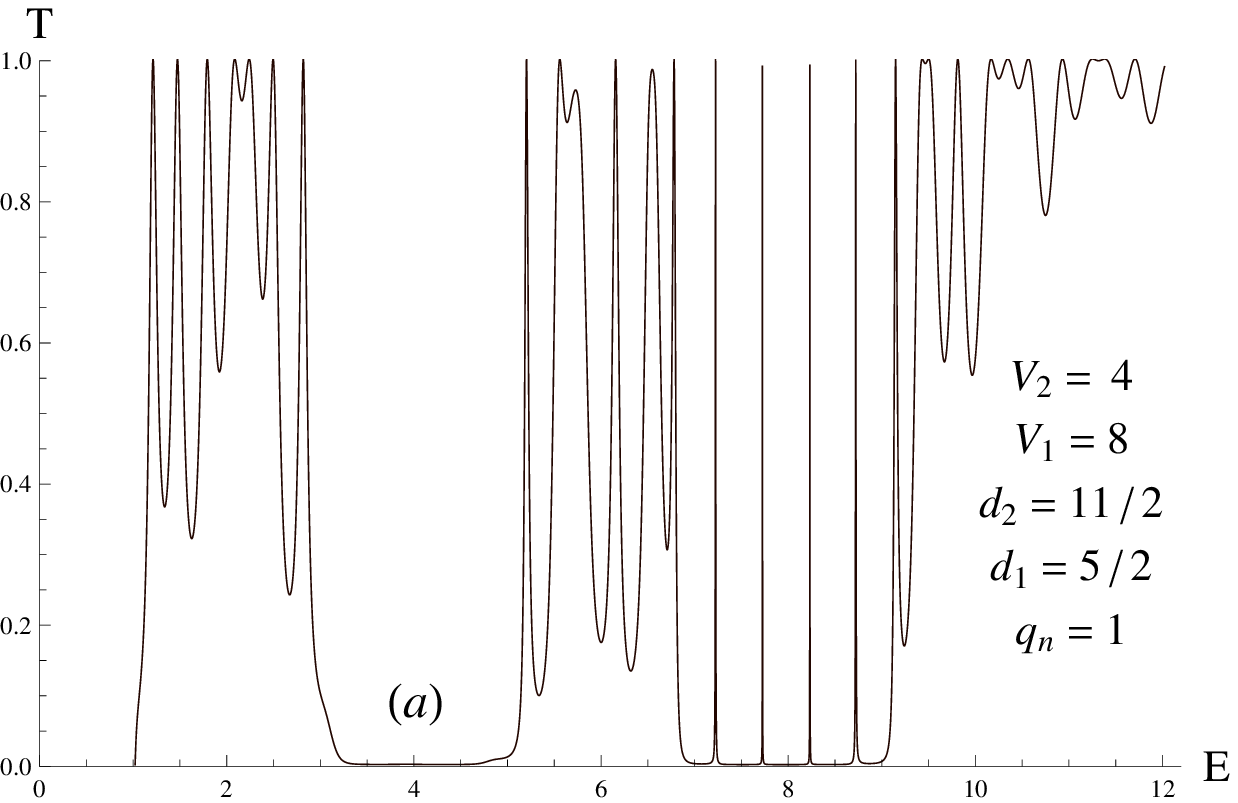}
 ~~~~~~~\includegraphics [width=7.5cm,keepaspectratio]
{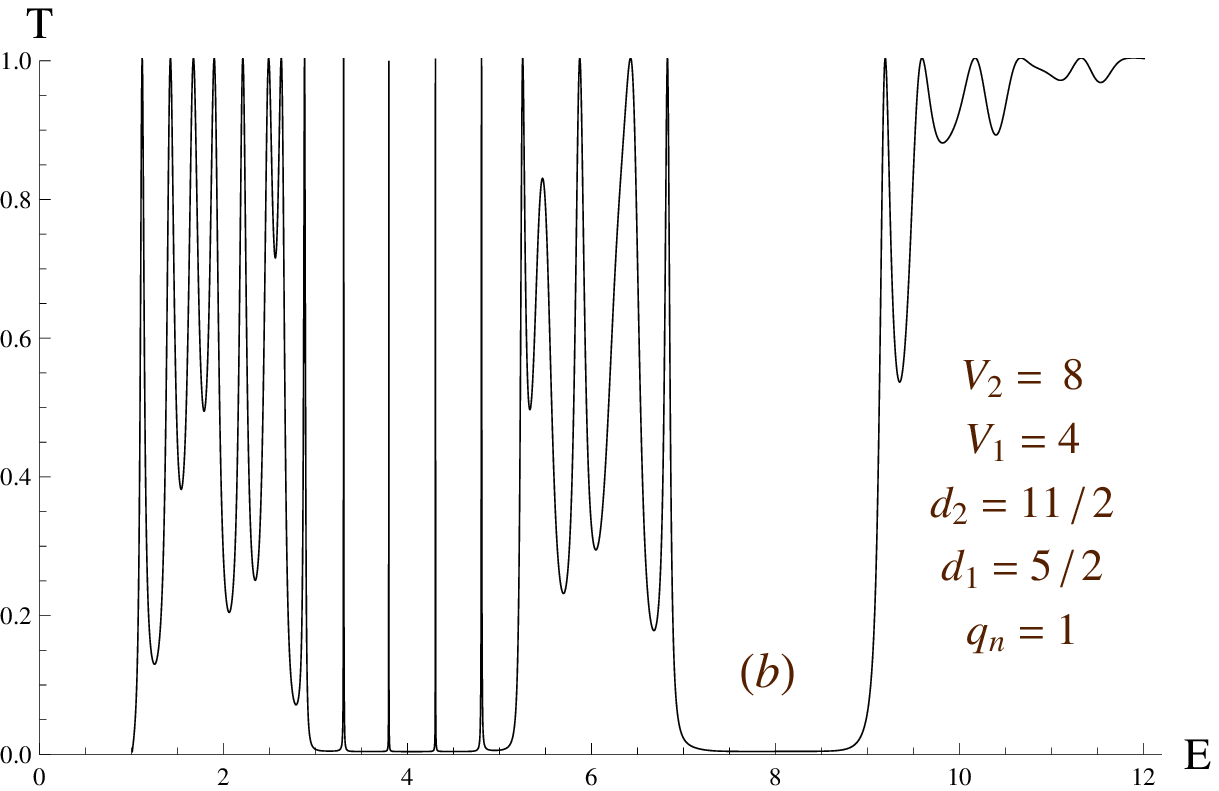}
\end{center}
{\sf{Figure 4: Transmission coefficient as a function of energy  for (a): $(V_0=0, \al<1)$ and  (b): $(V_0=0, \al >1)$.}}\\

\noindent In Figure 4a we distinguish  six different zones characterizing the behavior of the transmission coefficient.
These are

\begin{itemize}
\item The first is a forbidden zone  where $0\leq E \leq m^{*}$.
\item The second  is the lower Klein energy zone
characterized by resonances and $m^{*}\leq E \leq V_{2}- m^{*}$.
{Here we have full transmission at some specific energies despite the fact that the particle energy
is less than the height of the barrier. Note that this happens only if $V \geq 2m^{*}$. At these energies the wavefunction inside the barrier is oscillatory.}
\item The
third zone $V_{2}- m^{*}\leq E \leq V_{2}+ m^{*}$ is a window where
the transmission is zero, { the wavefunction is damped and transmission decays exponentially.}
\item The fourth $V_{2}+ m^{*} \leq E \leq
V_{1}- m^{*}$ is the higher Klein energy zone with transmission
resonances.
\item The fifth $V_{1}- m^{*} \leq E \leq V_{1}+ m^{*}$
is a window where the transmission is mostly zero but contains resonance
peaks {corresponding to the bound states associated with the double barrier.
Usually they are associated with the eigenvalues of the associated Hamiltonian or
poles of the associated Green function.}
\item The sixth zone $E \geq
 V_{1}+ m^{*}$ contains
oscillations, the transmission converges to unity at high energies similarly to the non-relativistic result.
\end{itemize}
Noting that Figure 4a for massless 2D  Dirac fermions coincides
with  transmission coefficient of massive  2D  Dirac fermions \cite{jellal2b}.
Contrary to the case $\al >1$ where $V_{2}$ becomes  larger than $V_{1}$
and therefore the results are modified, see Figure 4b. Indeed, compared to Figure 4a,
the behaviors of some zones are completely reversed like for instance the window zone.

One can also consider the interesting case $V_0 = V_2 = 0$, which is experimentally easier to
realize than $V_0= - \infty$ and $V_2 \neq 0$. Figure 5 shows the
behavior of the transmission coefficient, {which is completely changed
and shows only one Klein zone compared to Figure 4 which shows two Klein zones.}\\

\begin{center}
\includegraphics [width=7.5cm,keepaspectratio]
{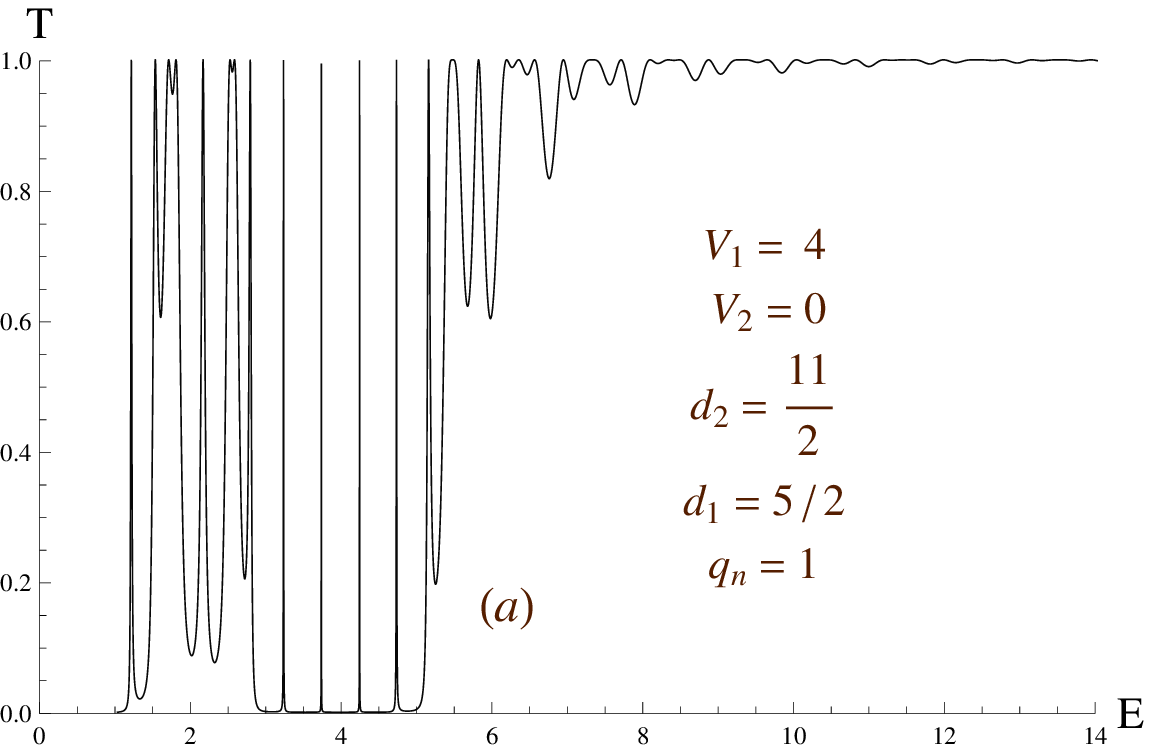}
 ~~~~~~~\includegraphics [width=7.5cm,keepaspectratio]
{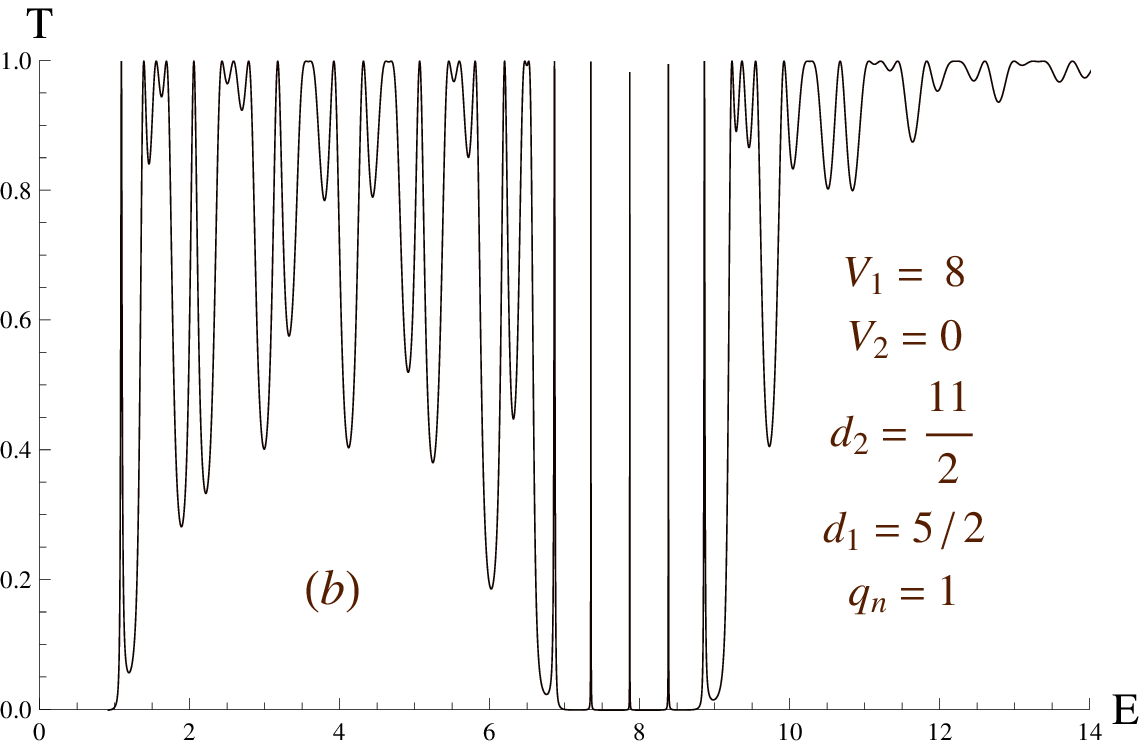}
\end{center}
{\sf{Figure 5: Transmission coefficient as a function of energy  for (a): $(V_0=V_2=0, V_1=4)$ and  (b): $(V_0=V_2=0, V_1=8)$.}}\\

At this stage, let us consider the case where $V_0\rightarrow - \infty$ and focus only on $\al >1$.
In this situation, the transmission amplitude for the $n$-th mode reduces to
\begin{equation} \label{GrindEQ__12_}
{t_{n} =\frac{2e^{2{i}[(k_{2} +k_{3} )d_{1} -(k_{1} -k_{2} )d_{2} ]}
(1+z_{2}^{2} )^{2} (1+z_{3}^{2} )}{A'+e^{4{i}(k_{3} d_{1} +k_{2} d_{2} )}
(1-z_{2} )^{2} (1+z_{2} z_{3} )^{2} +e^{4{i}k_{2} d_{1} } (1+z_{2} )^{2} (1+z_{2} z_{3} )^{2} } }
\end{equation}
 where the parameter $A'$ is given by
 \begin{eqnarray}
 A' &=& e^{4{i}k_{2} d_{2} } (z_{2} -1)^{2} (z_{2} -z_{3} )^{2} +e^{4{i}(k_{2} +k_{3} )d_{1} } (1+z_{2} )^{2} (z_{2} -z_{3} )^{2}  \non\\
 &&  +2e^{2{i}k_{2} (d_{1} +d_{2} )} (-1+e^{4{i}k_{3} d_{1} } )(1-z_{2}^{2} )(z_{2} -z_{3} )(1+z_{2} z_{3} ).
\end{eqnarray}
It is clear that the transmission coefficient depends on many parameters, such as the dimensions of
the graphene sheet $(w,l)$, the potential parameters $(d_1, d_2, V,\alpha)$, the propagation mode $n$ and the energy.
To study this behavior we proceed by varying the potential parameters
$w,l$ and $n$ while $E$ and $\al$ are fixed.
Figure 6 shows that the transmission coefficient contains two bowls, one centered at $V = E$ with resonant peaks
of unit amplitude and the other at $V =\frac{E}{\alpha}$ with vanishingly small transmission.
The geometry of the graphene system has different effects on the transmission coefficient. Indeed, when
$w$ increases the bowl width decrease and at certain value of $w$ the bowl depth starts to
decrease. However, if $l$ increases the number of oscillations of transmission increases as well.\\

\begin{center}
\includegraphics [width=7.5cm,keepaspectratio]
{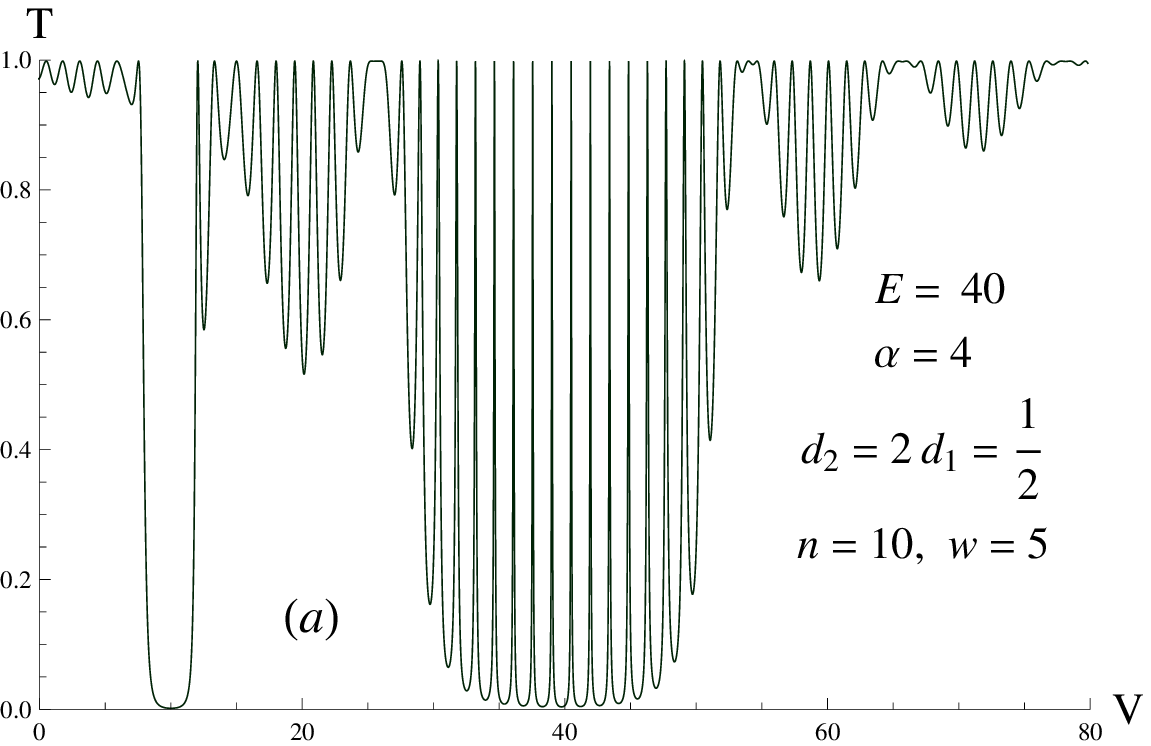}
~~~~~~~~~~~\includegraphics [width=7.5cm,keepaspectratio]
{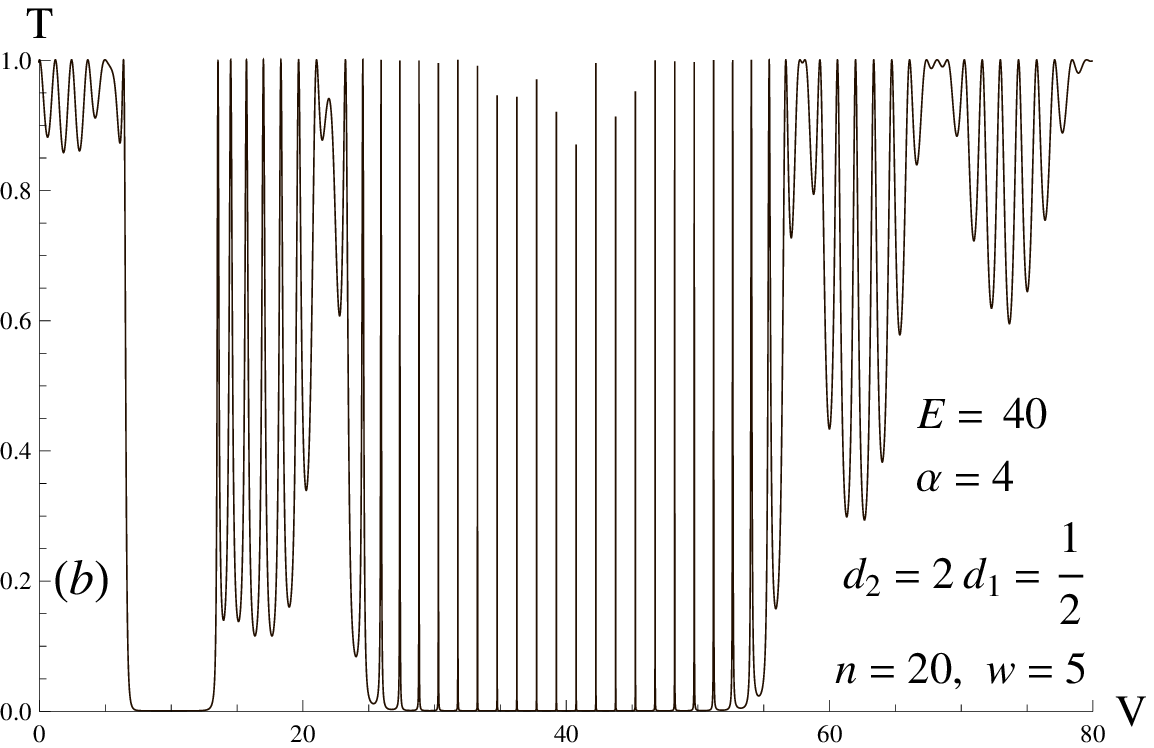}
\end{center}
\begin{center}
\includegraphics [width=7.5cm,keepaspectratio]
{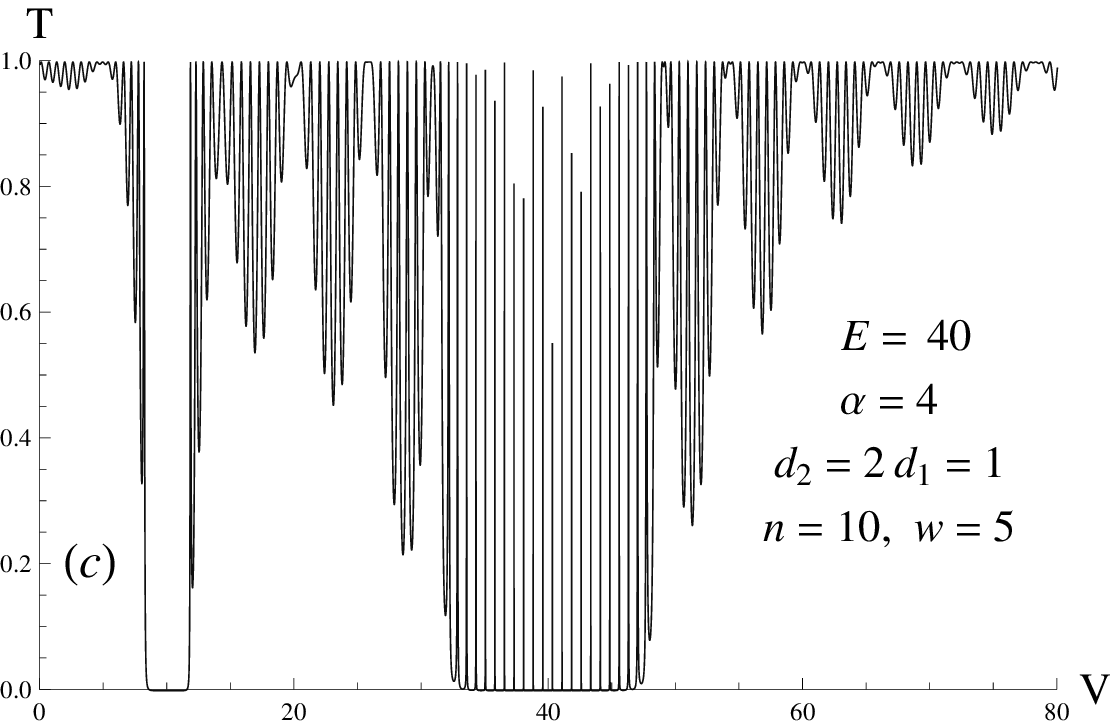}
~~~~~~~~~~~\includegraphics [width=7.5cm,keepaspectratio]
{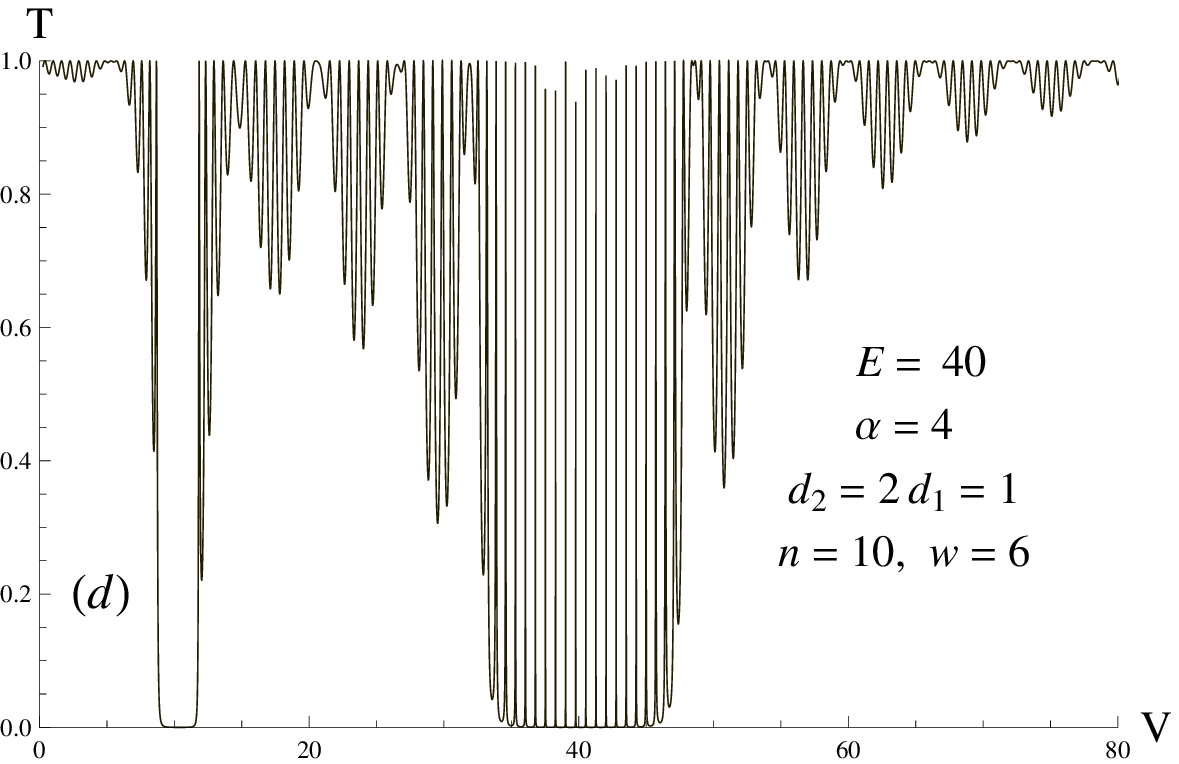}
\end{center}
%
%
{\sf{Figure 6: Illustration of the influence of each parameter of {(n,w,l)} on the transmission
coefficient as a function of potential $V$.}}

\section{Conductance and Fano factor}

To complete our study, we investigate two important physical quantities relevant to double barrier structure
in graphene, the conductance and the Fano factor. At zero temperature, they can be expressed,
respectively,  as
\begin{equation}\lb{GF}
  G=g_{0}\sum_{n=0}^{N-1}T_{n}, \qquad  F=\frac{
  {\sum_{n=0}^{N-1}T_{n}(1-T_{n})}}
  {
  {\sum_{n=0}^{N-1}T_{n}}}
\end{equation}
where $T_n=|t_n|^2$ is the transmission coefficient for $n$-th mode,
$g_{0} = \frac{4e^{2}}{h}$ and the factor $4$ accounts for the
spin and valley degeneracy. Taking into account the fact that  $w\gg l$,
we plot the conductance $\si=G\ \frac{l}{w}$ and Fano factor $F$ versus the
potential for different values of the energy. {This resulted in a similar graph
to that in~\cite{11} but shifted by an amount E along the potential axis. This is an
obvious result since the at zero temperature E represent the Fermi level.}\\
\begin{center}
\includegraphics [width=7.5cm,keepaspectratio]
{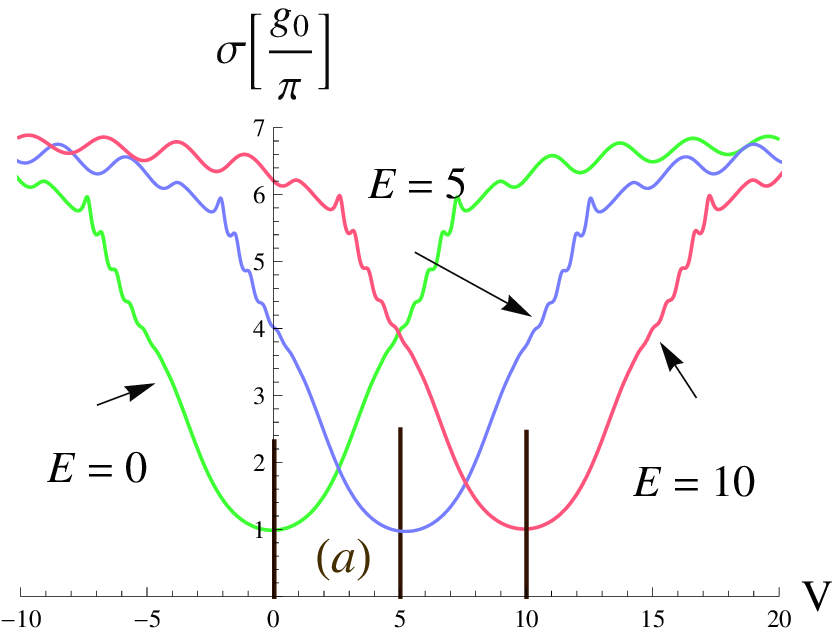}
~~~~~~~~~\includegraphics [width=7.5cm,keepaspectratio]
{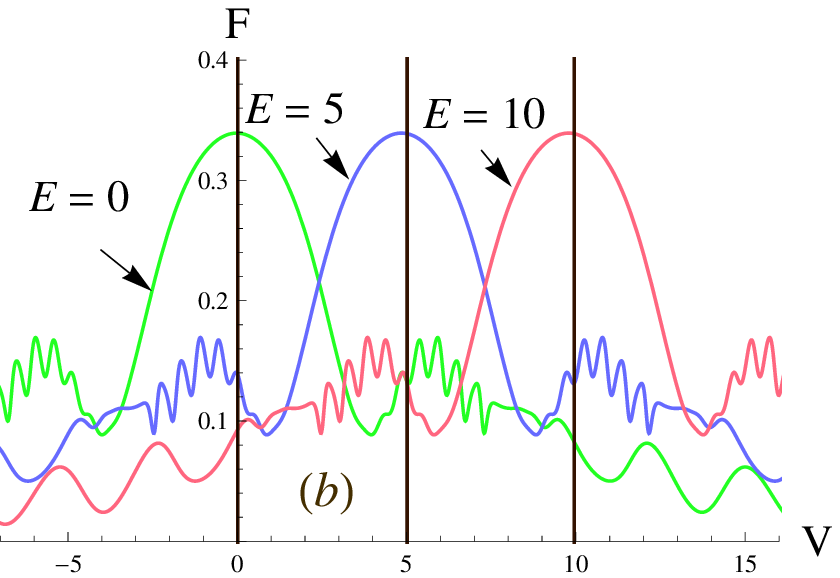}
\end{center}
{\sf{Figure 7: The conductance $\sigma$ and Fano factor $F$
as a function of the potential $V$ for different values of energy{, with $V_0=- \infty, \al=1$}.}}\\

\noindent {One can see that the conductance and the Fano factor keep the same generic behavior for all energies except that they have been
translated along potential axis, which means both of them are energy independent.
{In the Figure 8} we show the behavior of the conductance and the Fano factor as function of the potential V different values of $\alpha$.
From this figure it clear that $\sigma$ possess a minimum value
$\sigma=\frac{g_{0}}{\pi}$ but $F$ has a maximum value $F=\frac{1}{3}$ at zero energy independently of the value of $\alpha$. This result generalizes
that obtained in~\cite{11}, which was limited to the special case $\alpha = 1$}\\

\begin{center}
\includegraphics [width=7.5cm,keepaspectratio]
{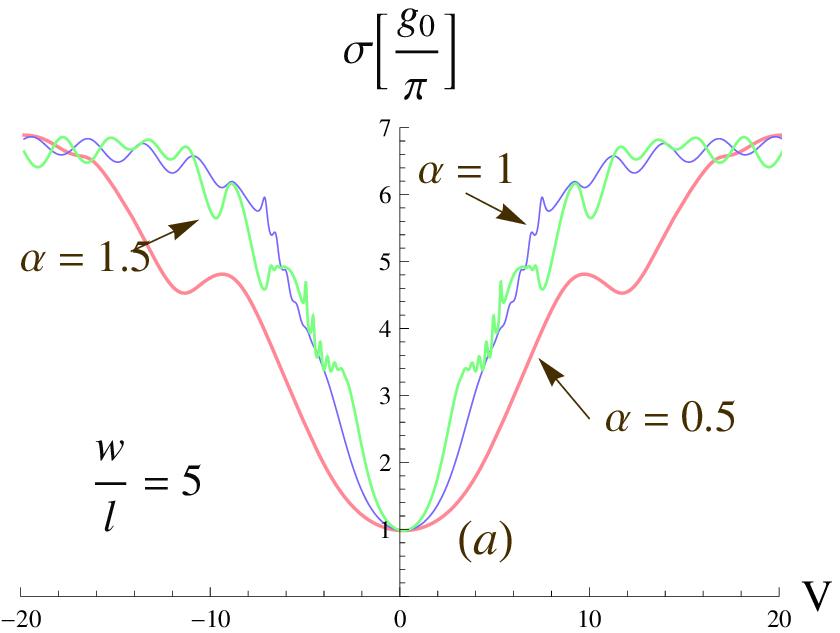}
~~~~~~~~~~~~\includegraphics [width=7.5cm,keepaspectratio]
{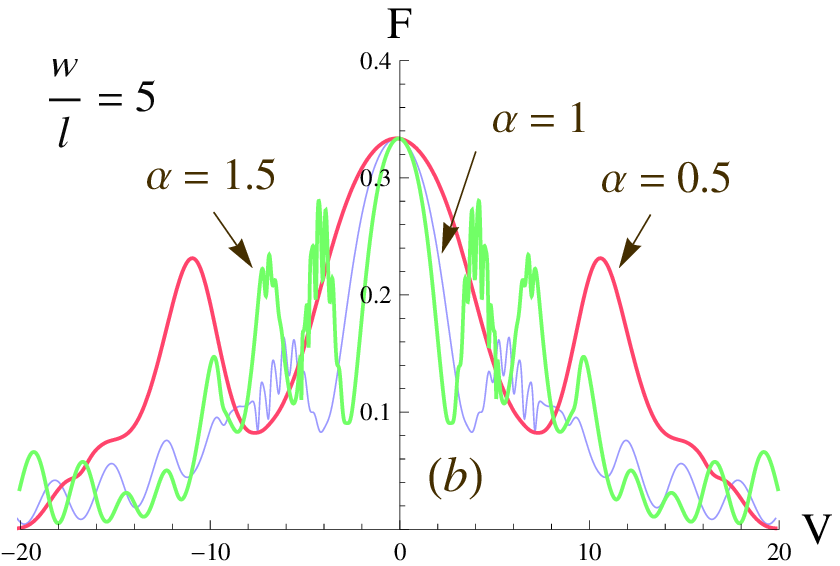}
\end{center}
{\sf{Figure 8: The conductance $\sigma$ and Fano factor $F$
as a function of $V$ at the Dirac point $E=0$, for different
values of $\alpha$ (blue curve for $\alpha=1$, red curve for
$0<\alpha<1$ and green curve for $\alpha>1 $), {with $V_0=-\infty$.}}}\\

\noindent From Figure 8 we can conclude that
for all values of $\alpha$,  we have the same conductance
minimum $\sigma=\frac{g_{0}}{\pi}$ and
Fano factor maximum  $F=\frac{1}{3}$ at the Dirac point $(E=0)$ but two extra satellite minima/maxima have been born.

It is worthwhile to investigate what happens if the energy is $E \neq 0$ and
$\alpha > 1$. This is described in Figure 9, which shows two conductance minima
at the points {${E}$}
and {$\frac{E}{\alpha}$} and similarly for the two maxima that occur for the Fano factor. It is interesting to note
that the first minimum in the conductance (Figure 9a) has no internal structure while the second one seems to have an oscillatory modulation
reminiscent of some interference phenomena. A similar observation can be made on Figure 9b for the Fano factor.\\

\begin{center}
\includegraphics [width=7.5cm,keepaspectratio]
{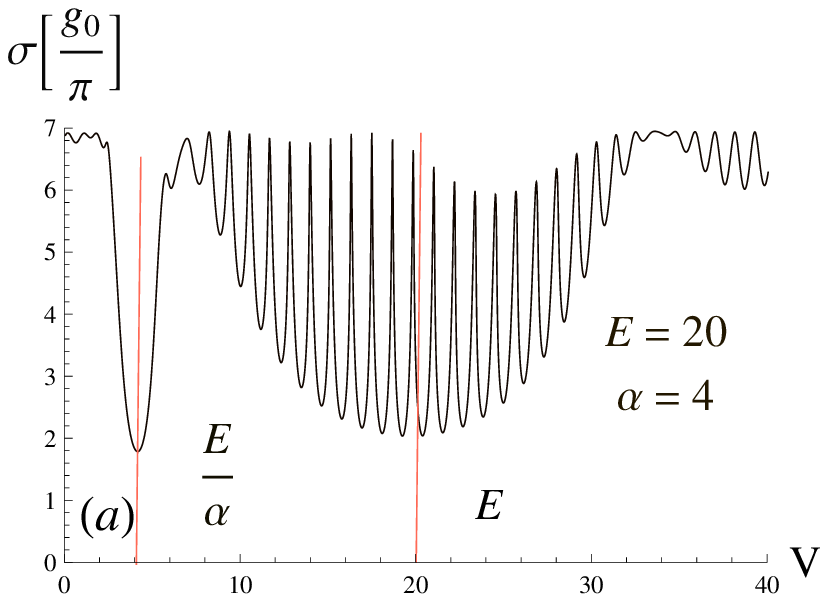}
~~~~~~~~~~~\includegraphics [width=7.5cm,keepaspectratio]
{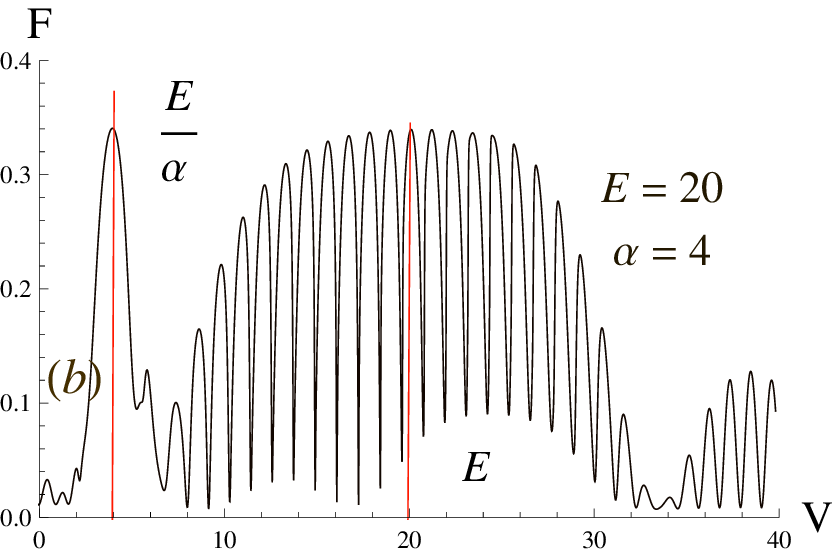}
\end{center}
{\sf{Figure 9: The conductance $\sigma$ and Fano factor $F$
in terms of the potential $V$ for $E>0$ and $\alpha=4$, {with $V_0=- \infty$}.}}\\


It is interesting to underline the behavior of
the conductance $\sigma$ and Fano factor $F$
in terms of  the inter-barrier distance $d_1$ for $V_2=0$.
The present case is plotted for different values of the potential
$V$ in Figure 10:\\

\begin{center}
\includegraphics [width=7.5cm,keepaspectratio]
{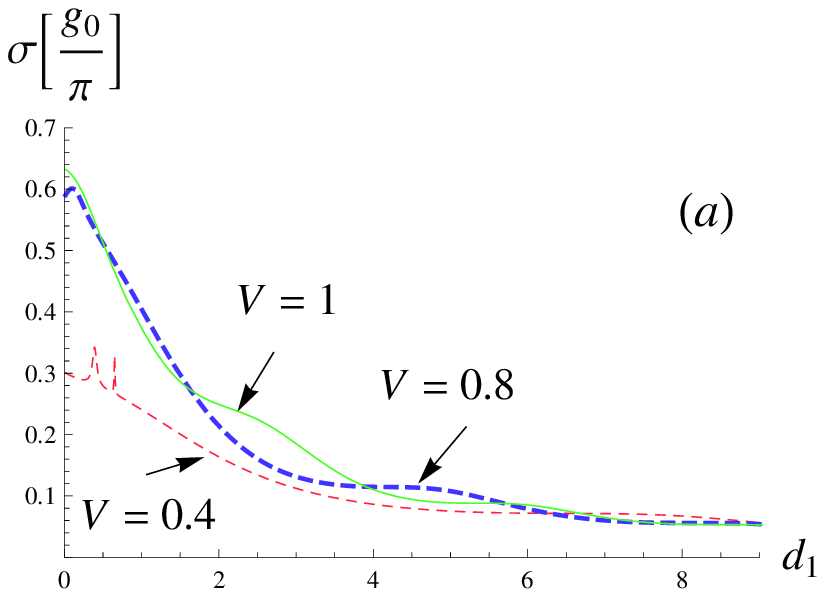}
~~~~~~~~~~~\includegraphics [width=7.5cm,keepaspectratio]
{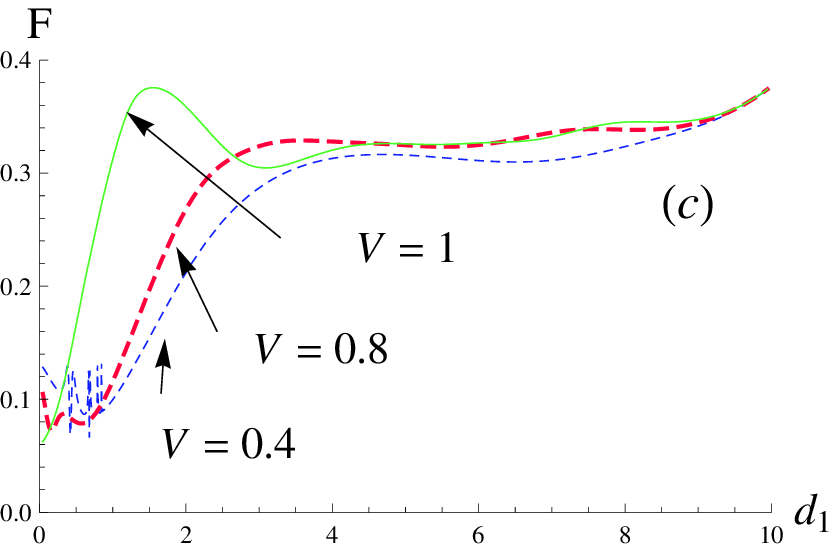}
\end{center}
\begin{center}
\includegraphics [width=7.5cm,keepaspectratio]
{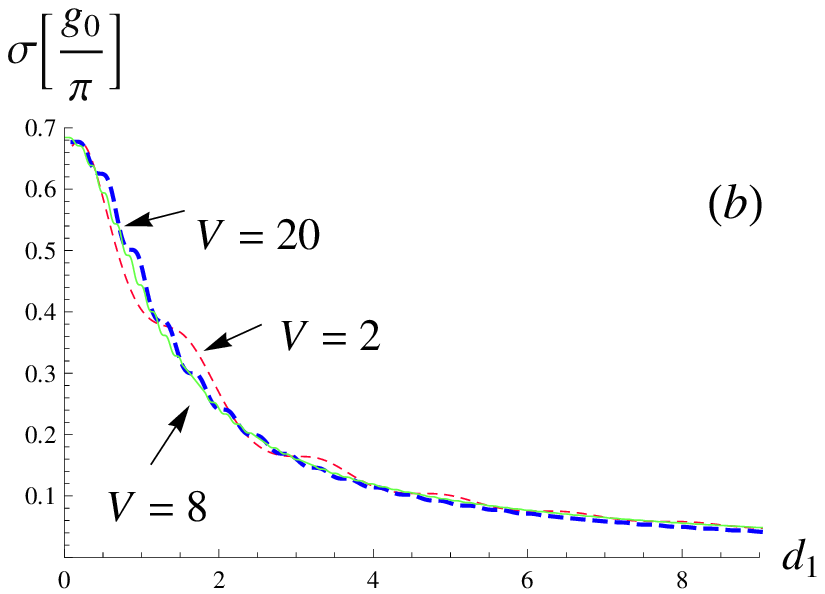}
~~~~~~~~~~~\includegraphics [width=7.5cm,keepaspectratio]
{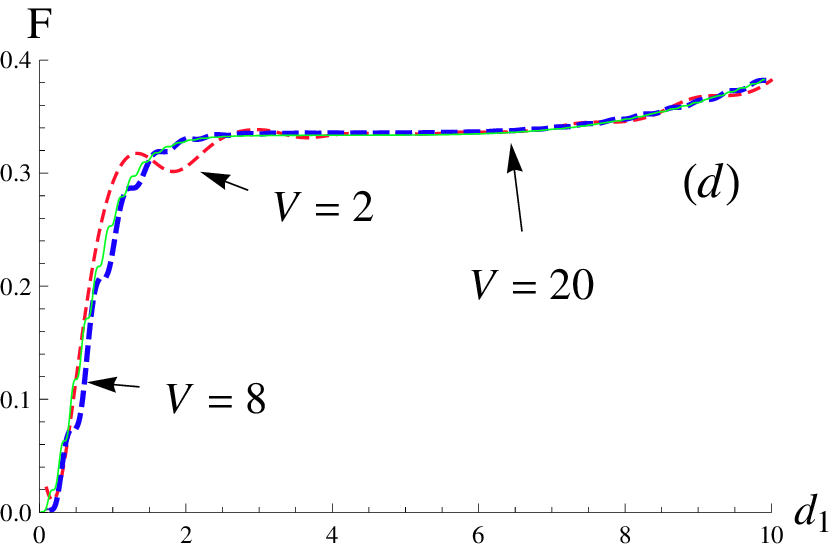}
\end{center}
{\sf{Figure 10: The conductance $\sigma$ and Fano factor $F$
versus the inter-barrier distance $d_1$ for $V_2=0$, {with $V_0=- \infty$}.}}\\ 

\noindent As shown in Figure 10, $G$ is decreasing as long as
 $d_1$ is increasing and  it is going to hold a constant value for large $d_1$.
However this is not the case for $F$, which is increasing
with $d_1$ up to certain value and becomes almost constant
for a large interval of $d_1$. As it is clearly shown in Figure 10b and 10d,
the behavior of $G$ and $F$ are not affected so much
by the variation of the potential $V>1$.

\section{Recovering the single barrier results}

Let us show how to recover one single barrier results presented in~\cite{11} from what
we obtained and discussed so far. We start by recalling that in \cite{11}
the authors  considered the single barrier case with the potentials
$V_{0}\rightarrow-\infty$ and $V_{1} =
V_{2}=V$. This allows us
to end up
with
the following restrictions on our parameters
\begin{eqnarray}
 && s_{1}=s_{5}= 1, \qquad
 \  k_{1}=k_{5}= \infty, \qquad
  z_{1}=z_{5}=1 \label{been1}\\
  && s_{2}=s_{3}= s_{4}, \qquad
  k_{2}=k_{3}=k_{4}, \qquad
  z_{2}=z_{3}=z_{4}. \label{been2}
\end{eqnarray}
Injecting the above results in
(\ref{GrindEQ__7_}) and  (\ref{GrindEQ__8_}), we get
the transmission amplitude
\begin{equation}
t_n=\frac{2 e^{-2{{i}}d_{2}(k_{1}-k_{2})}(1+z_{2}^{2})}{e^{4{{i}}d_{2}k_{2}}(1-z_{2})^{2}+(1+z_{2})^{2}}.
\end{equation}
Since the potential $V_{0}$ is negative then the wave vector $k_{1}$  is
purely real and consequently  $t_n$  can be reduced to the form
\begin{equation}
t_n=\frac{2(1+z_{2}^{2})}{e^{
{{i}}lk_{2}}(1-z_{2})^{2}+e^{
-{{i}}lk_{2}}(1+z_{2})^{2}}
\end{equation}
where $l=2d_{2}=4d_{1}$. Figure 11 shows the transmission behavior versus V for two values of propagation mode.\\

\begin{center}
\includegraphics [width=7.5cm,keepaspectratio]
{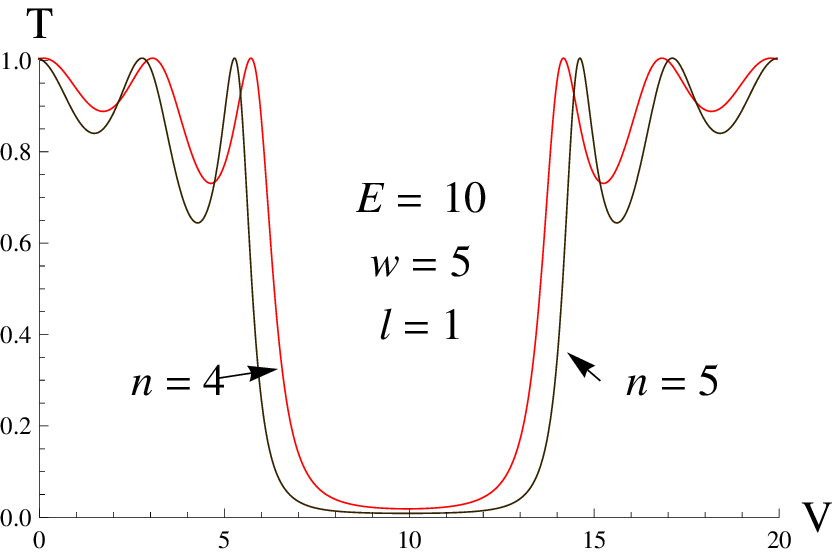}
\end{center}
{\sf{Figure 11: Transmissions coefficient as a function of the potential $V$ for two values of the propagation mode
$n=4$ and $n=5$, {with $V_0=-\infty$}.}}\\

\noindent It is clear that two transmission curves are symmetric with respect to
the point $V = E$ while an increase in the quantum number $n$ widen the bowl width.
On the other hand, setting $V_{0}\rightarrow-\infty$ and $V\rightarrow0$,
the transmission probability $T_{n}$ associated with the $n$-th mode at the
Fermi level is obtained from $T_{n} = |t_{n}|^{2}=|t|^{2}$ for $E =
0$. Under these circumstances our parameters are
given by
\begin{equation}
  k_{2}=i q_{n},  \qquad
  z_{2}\rightarrow\infty
\end{equation}
then our transmission amplitude takes a simple form
\begin{equation}
T_{n}=\left|\frac{2e^{lq_{n}}}{1+e^{2lq_{n}}}\right|^{2}=\cosh(lq_{n})^{-2}=\cosh\left[\frac{\pi
l}{{w}}\left(n+\frac{1}{2}\right)\right]^{-2}
\end{equation}
which coincides with that of \cite{11}.
Using (\ref{GF}), it is shown that
a minimum in the conductance and a maximum in the Fano factor \cite{11} occur at the Dirac point. In the limiting
case, i.e. $\frac{{w}}{l}\lga\infty$ (for a short and wide strip) at the Dirac point
$\si$ and $F$, respectively, reduce to
\begin{equation}
  \sigma\sim\frac{g_{0}}{\pi}, \qquad  F\sim\frac{1}{3}.
\end{equation}

Now let us consider a nonzero potential $V$
and evaluate the transmission coefficient at the Dirac point $E = 0$ to see
how the transmission obtained in \cite{11} will behave. Indeed, from our previous result
(\ref{zj2}) we can easily show that
\begin{equation}
  {\frac{(1\pm z_{2})^{2}}{1+z_{2}^{2}}=1\mp{V} k_{2}}.
\end{equation}
Using these results we obtain  transmission coefficient $T_{n}$ for the $n$-th mode as follows
 {\begin{equation}
T_{n}=\left|\frac{q_{2}}{k_{2}\cos(k_{2}l)+
i{V}\sin(k_{2}l)}\right|^{2}.
\end{equation}}
To illustrate this case and identify the difference with respect to
Figure 11, we plot the above result
for two particular values of $n$ in Figure 12:\\

\begin{center}
\includegraphics [width=7.5cm,keepaspectratio]
{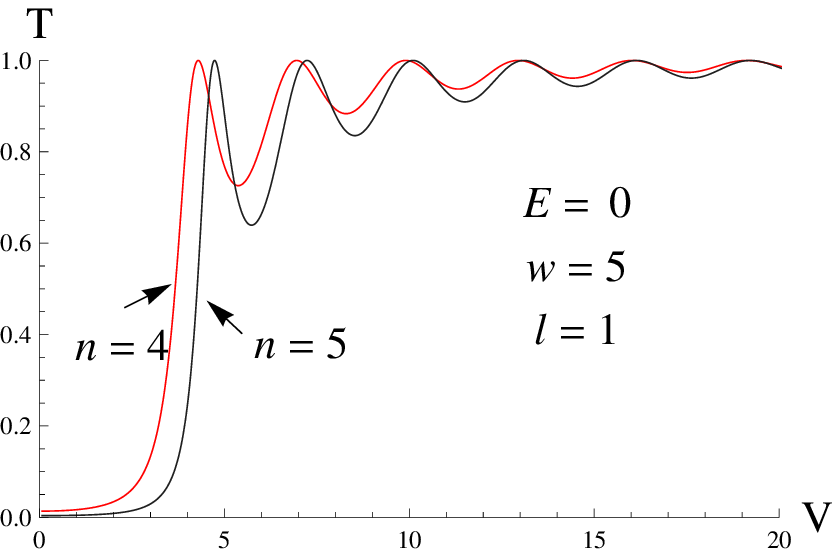}
\end{center}

{\sf{Figure 12: Transmissions coefficient as a function of $V$  at the Dirac point $(E=0)$
for $n=4$ and $n=5$, {with $V_0=-\infty$.}}}\\
%

\noindent One can see that both curves start from zero transmission
and oscillate to reach a total transmission, the valley gets wider as $n$ increases. This behavior
is similar to the 1D massive Dirac equation if one increase the mass of the Dirac Fermion, hence
establishing a strong relationship between the two systems.

\section{Conclusion}

We have generalized the results obtained
in~\cite{11} to double barrier case. More precisely,
we have analyzed transport properties through
double barrier structure in graphene. In the beginning, we have solved
the Dirac equation, for different regions forming the whole
system made of graphene, to get
the solutions of the  energy spectrum.
In the present analysis,
the infinite mass boundary
 condition was taken into account, which then resulted in a
 quantization of the wave number associated with the confining $y$-direction. This later
 is interpreted as a dynamical effective mass of Dirac fermions and compared to
 previously works.

Subsequently, the transmission of massless Dirac fermion through the double barrier structure
for various barrier parameters is studied.
 Continuity of the wavefunction at each interface along with the infinite mass boundary condition
 in the $y$-direction resulted
 in 
a system of eight algebraic equations for eight unknown coefficients.
 The detailed study of such system gave rise to a variety of interesting situations that were investigated.

 In fact, we have studied the effect of different potential parameters on the transmission,
 conductance and shot noise. An interesting situation
 arises when we have set the potential floor in the leads to zero, then our 2D problem reduces
 effectively to a 1D massive Dirac equation with
 an effective mass proportional to the quantized wave number along the transverse $y$-direction.
 Thus confinement along the $y$-direction generated
 effective masses for our fermions, which depend on the quantized wave number and
 its energy line spacing is proportional to the inverse
 of its width. Furthermore, we noticed that the minimal conductivity
and maximal Fano factor remain the same independently to the  ratio between the two
potentials $\left(\frac{V_2}{V_1}=\al\right)$.

Finally, we have shown how to recover from our model the single barrier results~\cite{11}.
This was done by considering the following limit  $V_0 = 0$ and $V_1=V_2=V$
in our general formulation. For the single barrier case with $V_0 = -\infty$ and $V_1=V_2=V$
we have recovered the result of reference~\cite{11}. This showed that our results are interesting
an more general.

\section*{Acknowledgments}
The generous support provided by the Saudi Center for Theoretical Physics (SCTP)
is highly appreciated by all Authors. We also acknowledge the support of  to KFUPM under project
RG1108-1-2. AJ also acknowledges partial support by King Faisal University.
The authors are indebted to the referee for his constructive comment.


\begin{thebibliography}{1}

\bibitem{1}  K. S. Novoselov, A. K. Geim, S. V. Morozov, D. Jiang, Y. Zhang, S. V. Dubonos,
I. V. Grigorieva and  A. A. Firsov, Science 306, 666 (2004).

\bibitem{2} C. Berger, Z. Song, T. Li, X. Li, A. Y. Ogbazghi, R. Feng, Z. Dai, A. N. Marchenkov,
E.H. Conrad, P.N. First and W. A. de Heer, J. Phys. Chem. B 108, 19912 (2004).

\bibitem{3} Y. Zhang, J. P. Small, M. E. S. Amori and P. Kim, Phys. Rev. Lett. 94, 176803 (2005).

\bibitem{4}  A. H. Castro Neto, F. Guinea and N. M. R. Peres, Phys. World 19, 33 (2006).

\bibitem{5} M. I. Katsnelson, Phys. Rev. B 74, 201401 (2006); Eur. Phys. J. B 51, 157 (2006).

\bibitem{6}  M. I.  Katsnelson and K. S. Novoselov, Solid State Commun. 143, 3 (2007).

\bibitem{7}  V. P. Gusynin, and S. G. Sharapov, Phys. Rev. Lett. 95, 146801 (2005).

\bibitem{8}  N. M. R. Peres, F. Guinea and A. H. Castro Neto, Phys. Rev. B 73, 125411 (2006).

\bibitem{11} J. Tworzydlo, B. Trauzettel, M. Titov, A. Rycerz and C. W. J. Beenakker, Phys. Rev. Lett. 96, 246802 (2006).

\bibitem{peeters1} M. Barbier, P. Vasilopoulos and F. M. Peeters, Phys. Rev. B 77, 205415 (2009).

\bibitem{peeters2} M. Ramezani Masir, P. Vasilopoulos and F. M. Peeters, Phys. Rev. B 82, 115417 (2010).

\bibitem{9}  P. R. Wallace, Phys. Rev. 71, 622 (1947).

\bibitem{10} D. P. DiVincenzo and E. J. Mele, Phys. Rev. B 29, 1685 (1984).

\bibitem{10'} J. Gonzalez, F. Guinea and M. Angeles H. Vozmediano, Phys. Rev. Lett. 69, 172 (1992).

\bibitem{12} M. V. Berry and R. J. Mondragon, Proc. R. Soc. Lond. A 412, 53 (1987).


\bibitem{kat} M. I. Katsnelson, K. S. Novoselov and A. K. Geim,  Nature Phys. 2, 620 (2006).


\bibitem{13} A. D. Alhaidari, A. Jellal, E. B. Choubabi and H. Bahlouli,
    {"Dynamical mass generation via space compactification in graphene"}, {\sf arXiv:1010.3437}.




\bibitem{Klein29} O. Klein, 
Z. Phys. 53, 157 (1929).

\bibitem{jellal2b} A. D. Alhaidari, H. Bahlouli and A. Jellal, {Relativistic Double Barrier Problem with Three Sub-Barrier Transmission Resonance Regions},
{\sf arXiv:1004.3892}.

\end{thebibliography}
\end{document}